%% file: RationalWL_16_ArXiv.tex
\documentclass[11pt,fleqn]{article}
\usepackage{amsmath,graphicx}

\input preamble.tex

\begin{document}
\twocolumn[

\Title{Three kinds of particles on a single rationally parameterized worldline}

\Aunames{Vladimir V. Kassandrov\auth{a,1} and Nina V. Markova\auth{b,2}}

\Addresses{
\addr a {Institute of Gravitation and Cosmology, Peoples' Friendship
	University of Russia, Moscow, Russia}
\addr b {Department of Applied Mathematics,  Peoples' Friendship
	University of Russia, Moscow, Russia}
	}

\Abstract
  {We consider the light cone (`retardation') equation (LCE) of an inertially moving observer and a  single worldline parameterized by arbitrary rational functions. Then a set of apparent copies, R- or C-particles, defined by the (real or complex conjugate) roots of the LCE will be detected by the observer. For any rational worldline the collective R-C dynamics is manifestly Lorentz-invariant and conservative; the latter property follows directly from the structure of Vieta formulas for the LCE roots. In particular, two Lorentz invariants, the square of total 4-momentum and total rest mass, are distinct and both integer-valued. Asymptotically, at  large values of the observer's proper time, one distinguishes three types of the LCE roots and associated R-C particles, with specific locations and evolutions; each of three kinds of particles can assemble into compact large groups - clusters. Throughout the paper, we make no use of differential equations of motion, field equations, etc.: the collective R-C dynamics is purely algebraic}
  
  ] 

\email 1 {vkassan@sci.pfu.edu.ru}
\email 2 {n.markova@mail.ru}

\section{Collective algebro-dynamics on a single `polynomial' worldline}
\label{sec1}

We intend to present some results in progress of the {\it `one electron Universe'}  concept formulated by Stueckelberg ~\cite{Stueckel}, Wheeler and Feynman~\cite{Feynman} and related to the so-called {\it unique worldline}. The latter can be either defined {\it implicitly}, by a set of algebraic equations containing the time parameter $t$~\cite{Khasan13,Mark14},  or in a familiar parametric way through consideration of the {\it light cone equation} (LCE) 
(equivalent to the well-known {\it retardation equation}) corresponding to an external observer~\cite{Khasan15}. In both cases, at some fixed value of $t$, one has a whole set of roots of the considered algebraic system which determine the positions and, consequently, temporal dynamics of the collection of identical particlelike formations. 

In the above cited papers it was shown that for an arbitrary worldline defined by {\it polynomial functions} (except a degenerate case of zero measure) the arising collective dynamics of the system of particles-roots is necessarily  {\it conservative}. This means that 
a complete set of conservation laws (for total momentum, angular momentum and the analogue of total energy) holds for the system of two kinds (R- or C-) of particlelike formations represented by real and complex conjugate roots of the generating set of equations, respectively.  It is especially interesting that all these laws follow solely from the structure of {\it Vieta formulas} for the whole system of roots, or from derivations of these formulas w.r.t. the time parameter.

In the case of implicitly defined polynomial worldline~\cite{Khasan13,Mark14} the considered {\it algebraic dynamics} is Galilei-invariant and can be compared with Newton's collective $N$-point dynamics. On the contrary, in the second case of a polynomial worldline implemented by the LCE of an {\it inertially moving observer} one obtains a full set of {\it Lorentz-covariant} conservation laws for the total set of R-C particles. Asymptotically, for rather great values of the observer's (proper) time $T$ 
one encounters, in addition, the effects of `self-quantization' of the admissible values of total rest mass  and `clusterization' of particle-roots. The possible meaning of the obtained {\it algebrodynamics}~\footnote{On the so-called {\it algebrodynamical} program see, e.g.,~\cite{AD,GR95,YadPhys} } for realistic relativistic physics requires further investigations. 

A detailed exposition of the above presented results can also be found in~\cite{dissovet}.             
Below we generalize our consideration of {\it polynomial} dynamics to the case of a worldline parameterized by arbitrary {\it rational} functions.

\section{Lorentz-invariant algebraic dynamics on a `rational' worldline}

In Euclidean 3D space $\bf E^3$, consider the LCE of an observer at rest, for simplicity, at the origin: 
\begin{equation}
(T-S(\tau))^2 -X(\tau)^2 -Y(\tau)^2 -Z(\tau)^2 =0,
\label{eq:LCE}
\end{equation}  
where $T$ represents the (proper) time of  the observer and $S:=X_0,\vec R:=\{X,Y,Z\}=\{X_a\},~ a=1,2,3$ are rational functions of the parameter $\tau$:
\begin{equation}
S(\tau):=\frac{S_{p+l}(\tau)}{S_p(\tau)},~~X_a(\tau):=\frac{X_{n+k}^{(a)}(\tau)}{D_n(\tau)} .
\label{eq:rational}
\end{equation}
In numerators and denominators one has arbitrary {\it mutually irreducible} polynomials in $\tau$ of corresponding degrees $p,n,p+l,n+k$, and it is assumed $l>k\ge 0$ to obtain nondegenerate polynomial parts (and an adequate asymptotic behavior, see Section 3). Explicitly selecting  polynomial and fractional parts for further consideration, one presents (\ref{eq:rational}) in the form:
\bearr
\label{polform}
X_0=f_0 \tau^l +g_0 \tau^{l-1} + h_0 \tau^{l-2} + \dots + \frac{S_{p-1}}{S_p},
\nnn \ \
X_a=f_a \tau^k +g_a \tau^{k-1}+h_a \tau^{k-2}+ \dots + \frac{X_{n-1}^{(a)}}{D_n}.
\ear  

 Under these assumptions, a finite set of $N=2p+2n+2l$ (real and complex conjugate together) roots $\{\tau_i(T)\}$ of the LCE (\ref{eq:LCE})  is sought for, which define the collective dynamics of $N$ pointlike objects associated with two types (R- or C, respectively) of particles. Note that any pair of C-particles can be visualized in $\bf E^3$ according to the common real parts of conjugate roots and thus represents a {\it composite} (C-) {\it particle of double mass}. 
 
At particular instants of the observer's time $T$ some two of the roots become multiple and then change their type (real to complex  conjugate or vice versa); the associated RC-particles merge and undergo mutual transitions. Such `events' can be interpreted as the process of annihilation/creation of a pair of R-particles (precisely, of a particle-antiparticle system)~\cite{Stueckel,Khasan13,Khasan15}. At the moments  of merging, the effective twistor  and electromagnetic fields (the latter of Lienard-Wiehert type) become singular on a null straight line connecting the points of observation and merging. The situation resembles the process of emission/adsorption of a lightlike carrier of interaction (a classical model of the photon).   
 
 Under Lorentz boosts, the observer's (proper) time $T$ and the worldline parameter $\tau$ remain invariant, while the coordinates transform in a canonical way. Since the LCE remains form-invariant,  effective collective kinematics (dynamics) of the RC-ensemble is also Lorentz-invariant, and the results can be carried over to the reference frame of any inertially moving observer. Note that all physical quantities should be here considered as functions of the {\it unique time} -- the proper time $T$ of the observer (instead of distinct  proper times of individual particles).   
    
Dispensing with the denominators in (\ref{eq:LCE}) and in the equation for the timelike coordinate $x_0$,~~$x_0-S(\tau)=0$, and eliminating from these two the parameter $\tau$ (say, making use of the corresponding {\it resultant} structure), one arrives at a polynomial equation
\begin{equation}
P_N (x_0,T) =0
\label{eq:timcoord}
\end{equation} 
of degree $N$ in $x_0$, in which the coefficients necessarily {\it depend polynomially} on $T$.  The same procedure also results in polynomial equations for any of the three spacelike coordinates of particles-roots $\{x_a\}$. 

Now, from the first two {\it Vieta formulas} (for the sums of roots  $x_0(T),x_a(T)$ and their squares) after necessary number of derivations w.r.t. $T$ one obtains two Lorentz-invariant conservation relations of the form ($\mu=0,1,2,3$):
\bearr
\label{eq:conserv}
\sum {\dot x}_\mu =constant =P_\mu, 
 \nnn \  \ 
 \sum {\dot x}_\mu {\dot x}^\mu + {\ddot x}_\mu x^\mu =const=M,
\ear
where `dot' designates $\partial/ \partial T$, and summing henceforth runs over all the $i=1,2,\dots,N$ roots of the generating LCE (\ref{eq:LCE})~\footnote{For simplicity, in what follows the summing index $i$ is not written out}. In the frame of reference of the observer at rest only the component $P_0$ is nonzero, whereas $\vec P:=\{P_a\}\equiv 0$ (the {\it center-of-mass} frame).  

In the case of a purely polynomial worldline~\cite{Khasan15} corresponding conserved quantities have been identified with the 4-vector $P_\mu$ of {\it total energy-momentum}  and the scalar of {\it total rest mass (rest energy)} $M$ of the RC-system, respectively. In this case (for $l>k$) it has been proved that these invariants identically {\it satisfy the fundamental relativistic energy-momentum relation} 
\begin{equation}   
 P_\mu P^\mu = M^2 >0
 \label{eq:energymom}
 \end{equation}
Moreover, `self-quantization' of the total rest mass values takes place for any polynomial worldline, $M$ being always positive, integer and equal to the full  number of particles-roots, $M=2l$.   

In the case of {\it rational} worldlines in question, the `self-quantization' effect still takes place. However, in this case it turns out that $P_0=2p+2l,~\vec P = 0$, and for the square of 4-momentum one gets
\begin{equation}\label{EnergyMomSq}
P^2=P_\mu P^\mu=(2p+2l)^2 >0, 
\end{equation}
whereas for the second invariant one obtains 
\begin{equation}\label{restmass}
M = 2p+2l-2n. 
\end{equation}
so that the inequality 
\begin{equation}\label{inequal}
P^2 > M^2
\end{equation}
holds for any R-C dynamical system. 

The above presented results can be proved by analyzing the structure of {\it resultants}, that is, determinants of corresponding Sylvester matrices. Quite a similar procedure has been used and exposed in details  in~\cite{Khasan15} for the case of  purely polynomial worldlines. Note that the second  invariant $M$ (identified previously with total rest mass) can be negative: this might be related to the contribution ($\sim 2n$) of negative {\it interaction energy}, a possible relativistic analogue of the potential energy.   

As for the values of total angular momentum $\vec M$, its conservation is affirmed by numerous computational experiments for worldlines with different $p,n,l>k$ and widely varying coefficients. Moreover, for some specific combinations of degrees one manages to establish phenomenological formulas for its values. For  example, in the case $l=2k-1$ (and arbitrary $p>1,n>1$) one obtains 
\begin{equation}\label{ang1}
\vec M = - \frac{2}{f_0} [~{\vec f} \times {\vec g}~],
\end{equation}
while for $l=2k-2$,
\begin{equation}\label{ang2}
\vec M = \frac{2g_0}{f_0^2} [~{\vec f} \times {\vec g}~] - \frac {4}{f_0} [~{\vec f} \times {\vec h}~],    
\end{equation}
where $\vec f:=\{f_a\}, \vec g:=\{g_a\}, \vec h:=\{h_a\}$ are 3-vectors composed from the  corresponding older coefficients in expressions for the parameterizing functions (\ref{polform}). Note that the purely fractional parts of these functions do not contribute to the angular moment values at all. For any $l\ge 2k$, the total angular momentum is also zero.

Finally, the other 3 components $\vec K:=\{K_a\} = \{M_{oa}\}$ of the relativistic tensor of total angular momentum 
$M_{[\mu\nu]}:=\sum (x_\mu \dot x_\nu - x_\nu \dot x_\mu)$ also preserve their values in time, according to numerous computational tests. Unfortunately, neither the proof of this fact nor the analytical formulas relating values of $\vec K$ with the coefficients of parameterizing functions (\ref{polform})  have not yet been obtained.

\section{Rationally parameterized worldline: three kinds of RC-particles}

Asymptotically, at large values of the observer's proper time $T$, three classes of roots can be distinguished, each defining its own set of RC-particlelike formations. Specifically, 
$2l$ roots (P-roots) are inherited from the structure of polynomial parts of generating functions and reproduce, on the whole, the asymptotic properties of the corresponding roots-particles  for the case of a purely polynomial worldline~\cite{Khasan15}. They are defined by the solutions for which one has $\vert \tau \vert \sim T>>1$, and the corresponding particles asymptotically join into pairs which then assemble into large groups -- {\it clusters} (under the condition of {\it mutually multiple} $l$ and $k$), undergoing mutual recession with retardation (for details  see~\cite{Khasan15,dissovet}). 

In the case of rational worldlines in question, there exists, however, another class (S) of pairs of roots-particles corresponding to rather small values $\vert \tau \vert \sim 1$. The constituent particles asymptotically approach one another and one of the $p$ zeros of the polynomial denominator $S_p(\tau)$ of the timelike coordinate function $S(\tau)$. There are exactly $2p$ roots-particles of this kind: they are located at small distances $\vert x_a \vert \sim 1$ and move with rather small velocities $v\sim 1/T^2 <<1$. Finally, the third kind (D) of roots-particles ($2n$ in number) relates to the zeros of the `spatial' polynomial denominator $D_n(\tau)$. The arising pairs  are located at great distances and move with ultra-relativistic velocities $v\sim 1$. 

It is especially interesting when the polynomials - denominators  $S_p(\tau)$ and/or $D_n(\tau)$ possess {\it multiple} roots (of possibly great multiplicity). Then, apart from the clusters related to P-class particles, one encounters two other sorts of clusters which are due to the asymptotic generation of roots close to those of denominators $S_p(\tau)$ and/or $D_n(\tau)$. Particles of the first (S) class compose big groups located near the observer and possessing nonrelativistic velocities. Clusters of he second (D) class consist of distant particles moving with velocities asymptotically approaching the speed of light (contrary to P-clusters for which velocities asymptotically tend to zero).  More details of the general situation will be illustrated on a typical example presented below.  

\section{Example and discussion}

As an example of a rationally parameterized worldline, let us choose, in a random way, polynomial parts with (mutually multiple) $l=16, k=9$ and supplement these by fractional parts with, say, $p=6, n=6$. Besides, let the denominator $S_p(\tau)$ possess a root $\tau=2$ of multiplicity 4 and $\tau=5$ of multiplicity 2, while the denominator $D_n(\tau)$ has a root $\tau =-1$ of multiplicity 3 and $\tau=-7$ of multiplicity 3. Specifically, let us take the coefficients of parameterizing functions (\ref{polform}) as 
\bearr
\label{paramexam}
f_0=1,~g_0=-3,~h_0=-4;~~\vec f = \{1,2,-3\}, 
\nnn \ \
\vec g =\{-11,17,13\},~\vec h = \{-3,7,-9\};  
\ear
and, in the numerators of the fractional parts,
\bearr 
\label{frac}
S_{p-1}=-4\tau^5+\dots,~~X_{n-1}=-11\tau^5+\dots, 
\nnn \ \ 
Y_{n-1}=5\tau^5+\dots,~~Z_{n-1}=14\tau^5+\dots,
\ear
while the denominators are of the form 
\begin{equation}\label{denom}
S_p=(\tau-2)^4 (\tau-5)^2,~D_n=(\tau+1)^3 (\tau+9)^3.
\end{equation}
(the omitted terms are of no need for the reader).

Reducing the l.h.s. of the LCE and the equation $x_0 - S(\tau)=0$ to common denominators and calculating the resultant of the two polynomials represented by the corresponding numerators, we obtain the equation 
\begin{equation}\label{RS}
Ax_0^{56}+BTx_0^{55}+CT^2 x_0^{54}+\dots =0,
\end{equation} 
($A,B,C$ are very large integers whose values are not important here), 
from which we get the whole number of particles-roots $N=56=2p+2l+2n$ and, making use of Vieta formulas, the conservation laws for the derivatives of the timelike coordinate ${\dot x}_0:=dS/dT$,
\bearr
\label{conservt}
P_0=-B/A = 44,~~W_0:=\sum {\dot x}_0^2+ x_0 {\ddot x}_0 = 
\nnn \ \
    (B/A)^2-2C/A=44 = P_0 \equiv 2p+2l. 
\ear
Using a similar procedure for spatial coordinates, we get at the end
\bearr
\label{conservx}
P_a = 0,~~W_a:=\sum {\dot x}_a^2+ x_a {\ddot x}_a=\{ 
\frac{209828990102272}{852207569447017},
\nnn \ \
\frac{231793932340096}{852207569447017},~\frac{581026160921836}{852207569447017}\},
\ear
so that for Lorentz invariants (\ref{EnergyMomSq}),(\ref{restmass}) one has 
\bearr
\label{invar}
P^2=P_\mu P^\mu = 1936=44^2 \equiv (2p+2l)^2,~M =  
\nnn \ \
W_0 - W_1 - W_2 - W_3 =32 \equiv 2p+2l-2n.
\ear
Now, for the total angular momentum one obtains
\begin{equation}\label{totang}
\vec M = \{-474, -192, -286\},
\end{equation}
 in full agreement with Eq.(\ref{ang2}). Finally, the remaining components $\vec K$ of the angular momentum tensor are rational numbers of order $10^7$. 
    
At large values of the observer's time, say, at $T=10^{14}$, there are 7 simple~\footnote{Multiparticle clusters require a great multiplicity of $l$ and $k$; corresponding situation is illustrated in~\cite{Khasan15,dissovet}}  groups - clusters consisting of 2 composite C-particles ($7\times 4 =28$ roots) and 2 pairs of R-particles ($2\times 2 = 4$ roots) corresponding altogether to $2l=32$ roots of P-class for which $\vert \tau \vert \sim T >> 1$, and the LCE can be approximated by the reduced form $(T-S(\tau))^2\sim 0$. One particle in any pair  corresponds to retarded ($T-\Re{(S)}>0$) while the other to advanced ($T-\Re{(S)}<0$) solutions to the LCE. All these are located at a distance $R\sim  T^{k/l}=T^{9/16}$ (see~\cite{Khasan15})  from the observer and slowly recess, whereas mutual distances slowly decrease (as compared with $R$). For more details  see~\cite{Khasan15,dissovet}.

Further, one has 8 roots of S-class forming the cluster of two R-pairs (4 roots) and one C-pair (4 roots) of particles, and 4 roots of S-class consisting of a pair of C-particles, $2p=12$ of S-roots altogether. They correspond to the values  of $\tau$ very close to the zeros ($2$ and $5$) of the denominator $S_p$. Both clusters are rather close to the observer ($R\sim 10^4$ or $10^6$, respectively) as well as the constituent RC-particles which move with negligibly small velocities ($v\sim 10^{-13}$ and $10^{-7}$). Rather a large number of roots-particles (8) in the first cluster correlates with the multiplicity 4 of the root $\tau=2$ of the denominator $S_p$. One half of the particles-roots corresponds  again to  retarded and the other half to advanced solutions. 
    
Finally, there are two groups of 6 roots of D-class corresponding to zeros $-9$ and $-1$ of the denominator $D_n$. However, due to the odd degree of multiplicity of both zeros, each group of roots gives rise to 2 clusters of 3 particles disposed at {\it opposite directions of the celestial sphere}. The typical distances of all clusters are $R\sim T$, and {\it velocities of all particles are  very close to the speed of light}, $v\sim 1$. Particles in the first two clusters correspond to advanced solutions of the LCE; they all approach the observer (from opposite directions). Conversely, particles in other two clusters are retarded and recess from the observer (in opposite directions): these structures bear most  resemblance of `cosmological' objects. 

\vskip2mm

To conclude, we have discovered some remarkable properties of algebraic structures related to the specific form of the LCE and rationally parameterized worldlines. These properties widen the previously obtained~\cite{Khasan15} possibilities for description of N-point collective relativistic dynamics on the basis of `polynomial' worldlines. As compared to the polynomial case, the algebraic dynamics on a rational worldline (asymptotically) involves 3 R- and 3 C- distinct kinds of particlelike formations, with rather diverse and peculiar kinematic behavior. Specifically, at late times one encounters distant `objects' recessing with negative acceleration, distant objects moving with ultra-relativistic velocities and slowly moving nearby objects. 

In general, all these are joined into large clusters consisting of a lot of constituent particles-roots.  Contrary to  the `polynomial' case (in which recession takes place along one or two privileged 3-directions~\cite{Khasan15}), particles of S- and D-classes can at late times be distributed {\it isotropically}. As a whole, the pattern of evolution partly resembles Milne's {\it kinematic relativity}~\cite{Miln}; comparison of the two models will be considered elsewhere.

 However, of utmost interest from a physical viewpoint is certainly the Lorentz-invariant and  conservative nature of the collective RC-dynamics inherited from the polynomial case, as well as the `self-quantization' property of the two principal Lorentz invariants. Remarkably, here the square of 4-momentum is always greater than the total rest mass (see (\ref{inequal})), so that the famous mass-shell condition in SR (valid, in fact, only for free or locally colliding particles) is generalized here to the case of an arbitrary {\it interacting} N-particle system. 
 
Of course, one can hardly believe that the elaborated construction has an explicit relation to the real physical world: more probably, it can be treated as a sort of a `toy model'. On the other hand, the fact that all the above presented properties follow from only purely mathematical considerations, free of any phenomenological speculations, seems quite remarkable and unexpected. The existence of such structures cannot be ignored: they certainly deserve further mathematical investigations and proper physical interpretations.

\vskip2mm
The authors are grateful to Joseph A. Rizcallah for fruitful discussions and comments.

\end{document}

%% file: preamble.tex
\usepackage{amsfonts,amssymb,cite}
\usepackage{epsf,graphicx}

\def\noi{\noindent}

\newcommand{\Title}[1]{\noi {{\Large\bf #1}}\\[1ex]}

\def\Aunames#1{\noi{\bf #1}}
\def\auth#1{${}^{#1}$}
\def\Addresses#1{\medskip\noi \protect
	\begin{description}\itemsep -3pt {\it #1} \end{description}}
\def\addr#1#2{\item[${}^{#1}$]{\it #2}}

\newcommand{\Abstract}[1]{\vskip 2mm \begin{center}
        \parbox{16.4cm}{\small\noi #1} \end{center}\medskip}
\newcommand{\PACS}[1]{\begin{center}{\small PACS: #1}\end{center}}

\def\email#1#2{\footnotetext[#1]{e-mail: #2}\addtocounter{footnote}{1}}

\def\nqq{\hspace*{-2em}}

\def\Jl#1#2{#1 {\bf #2},\ }

\def\ApJ#1 {\Jl{Astroph. J.}{#1}}
\def\CQG#1 {\Jl{Class. Quantum Grav.}{#1}}
\def\DAN#1 {\Jl{Dokl. AN SSSR}{#1}}
\def\GC#1 {\Jl{Grav. Cosmol.}{#1}}
\def\GRG#1 {\Jl{Gen. Rel. Grav.}{#1}}
\def\JETF#1 {\Jl{Zh. Eksp. Teor. Fiz.}{#1}}
\def\JETP#1 {\Jl{Sov. Phys. JETP}{#1}}
\def\JHEP#1 {\Jl{JHEP}{#1}}
\def\JMP#1 {\Jl{J. Math. Phys.}{#1}}
\def\NPB#1 {\Jl{Nucl. Phys. B}{#1}}
\def\NP#1 {\Jl{Nucl. Phys.}{#1}}
\def\PLA#1 {\Jl{Phys. Lett. A}{#1}}
\def\PLB#1 {\Jl{Phys. Lett. B}{#1}}
\def\PRD#1 {\Jl{Phys. Rev. D}{#1}}
\def\PRL#1 {\Jl{Phys. Rev. Lett.}{#1}}

\def\lal{&&\nqq {}}

\def\beq{\begin{equation}}
\def\eeq{\end{equation}}
\def\bear{\begin{eqnarray}}
\def\bearr{\begin{eqnarray} \lal}
\def\ear{\end{eqnarray}}
\def\earn{\nonumber \end{eqnarray}}

\def\nnn{\nonumber\\ \lal }

